\documentclass[review]{elsarticle}

\usepackage{lineno,hyperref}

\journal{NIM}

\begin{document}

\begin{frontmatter}

\title{Beam tuning and bunch length measurement in the bunch compression operation at the cERL}

\author[kekadd,sokenadd]{Y.Honda}
\author[kekadd,sokenadd]{M.Shimada}
\author[kekadd,sokenadd]{T.Miyajima}
\author[sokenadd]{T. Hotei}
\author[kekadd,sokenadd]{N.Nakamura} 
\author[kekadd,sokenadd]{R.Kato}
\author[kekadd,sokenadd]{T.Obina}
\author[kekadd,sokenadd]{R.Takai}
\author[kekadd,sokenadd]{K.Harada}
\author[kekadd]{A.Ueda}

\address[kekadd]{High Energy Accelerator Research Organization (KEK), 1-1 Oho, Tsukuba, Ibaraki, Japan}

\address[sokenadd]{The Graduate University for Advanced Studies (SOKENDAI), 1-1 Oho, Tsukuba, Ibaraki, Japan}



\begin{abstract}
Realization of a short bunch beam 
by manipulating the longitudinal phase space distribution
with a finite longitudinal dispersion following an off-crest acceleration 
is a widely used technique.
The technique was applied in a compact test accelerator of an energy-recovery linac scheme
for compressing the bunch length at the return loop.
A diagnostic system utilizing coherent transition radiation 
was developed for the beam tuning and for estimating the bunch length.
By scanning the beam parameters,
we experimentally found the best condition for the bunch compression.
The RMS bunch length of 250$\pm$50 fs was obtained
at a bunch charge of 2 pC.
This result confirmed the design 
and the tuning procedure of the bunch compression operation for
the future energy-recovery linac (ERL).
\end{abstract}


\end{frontmatter}


\section{Introduction}

The short pulse electron bunch produced by an accelerator
is applied in a variety of applications.
For example,
the electron pulse itself \cite{UedYang}
or 
the short pulse X-ray produced by synchrotron radiation from the electrons
is useful for time-resolved observation of fast phenomena.
The short and intense X-ray pulse
recently available in free electron lasers (FEL) \cite{Lcls}\cite{Sacla}
is the most advanced source in the field.
The coherent radiation in the terahertz (THz) frequency range
produced by a structure in a bunch shorter than the wavelength 
is expected to be useful in the field of molecular structure analysis.
Such an accelerator-driven THz source
achieves the highest power in the frequency range 
among all the schemes \cite{jlabcsr}\cite{FlashCTRterahertzsource}.

Compared with storage-ring-based accelerators,
where the bunch length is determined by the equilibrium condition of the ring,
linac-based single-path accelerators have an advantage in realizing a short bunch length.
In a typical linac system,
the beam at the electron gun and the low-energy section
is designed to have a relatively long bunch duration to suppress the space-charge force.
It is compressed in several stages of the bunch compression system
designed along the acceleration to achieve a short bunch at the end.
The performance of the bunch compression system 
is key to achieving a short bunch in a linac system.

A conventional linac,
however, has a limitation with respect to the average beam power
because the operational pulse repetition is limited due to the energy efficiency.
An energy-recovery linac (ERL) scheme with superconducting rf cavities
has opened the possibility of high-beam-power operation
with a linac-based accelerator \cite{erlreview}. 
The short bunched beam at a high average power,
which is available only from an ERL, is necessary
for future light sources such as a high power FEL.

In KEK,
we have constructed a test ERL accelerator, the compact ERL (cERL \cite{Sakanakun}),
for the development of a future larger-scale ERL facility.
It contains many of the key components of the ERL accelerator.
The validation of the design of hardware components and problems in beam dynamics,
including the procedure for beam commissioning,
will be performed during operation of the test accelerator.
Because the beam optics design of the cERL
is considered to be scalable to a larger machine,
the common problems in the beam tuning can be tested.
In particular, the bunch compression scheme utilizing the arc section
is an important test item in the cERL.

In order to tune the beam for bunch compression,
a bunch length diagnostics is necessary.
Many types of bunch length monitors have been studied 
in various accelerator laboratories.
The most direct and reliable one is the RF deflector system.
It uses the time-varying field of an RF cavity 
to project the time structure of the bunch into a transverse space.
However, for an ERL accelerator targeting operation of a high current beam,
installation of an RF structure is not realistic.
An electro-optical method
that samples the electric field moving with the bunch
can also be regarded as a direct measurement.
However the strength of the electric field 
is too low in our case of a low bunch charge.
An indirect method to observe the spectrum of a coherent radiation
has been widely used.
Although some ambiguities remain
in the bunch shape
due to the limitation in the detection range of the spectrum,
it is a handy method to use. 
Several coherent radiation mechanisms can be used for the measurement.
Coherent synchrotron radiation (CSR) from 
a bunched electrons in a bending magnet
can be used without disturbing the beam, 
and hence it can be applied in a high-current accelerator.
However, 
as a bunch length monitor,
it has ambiguities in the source point and source size
because the source point of CSR is in a bending magnet.
Coherent transition radiation (CTR) produced
by inserting a target at a straight section
is suitable for our beam diagnostics.
We have developed a bunch length monitor utilizing CTR.

In this paper,
we report the first bunch compression study
to confirm the beam tuning procedure of the cERL
conducted at a relatively low bunch charge
neglecting collective effects.

\section{Principle}

\subsection{Principle of bunch compression system}
\label{sec:principle}

\begin{figure}[htb]
	\begin{center}
	 \includegraphics[width= 0.8\linewidth]{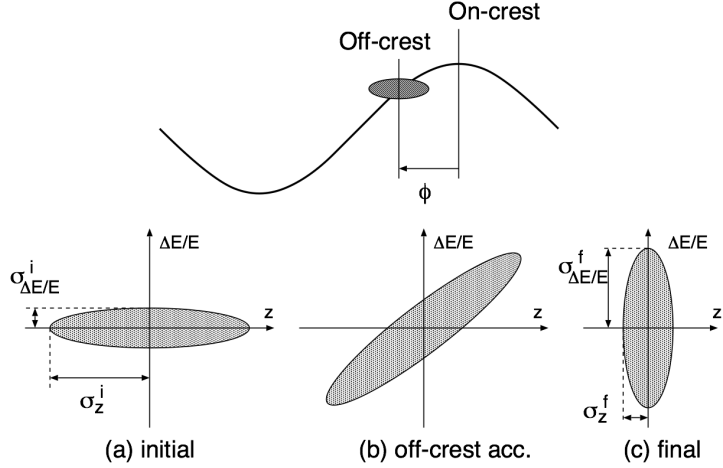}
	  \caption{
Phase space in the bunch compression.
The timing of the bunch is set at a shifted phase from the maximum acceleration.
The initial bunch (a) having long duration but small energy spread 
is converted to a bunch having (c) short duration but large energy spread.
}
	\label{fig:phasespace}
\end{center}
\end{figure}

The bunch length compression
is achieved by manipulating the bunch in the longitudinal phase space
using an off-crest acceleration and a non-zero longitudinal dispersion
\cite{handbook} \cite{ipac_shimada} \cite{ipac_nakamura} \cite{linac_piot}.
Here, we consider a situation where
the initial beam energy is $E_i$,
which is further accelerated up to $E_f$ using an RF accelerator set to be off-crest.
The development of the beam volume in the phase space
is explained in Fig.\ref{fig:phasespace}.
(a) shows the initial distribution.
We express the RMS spread in the longitudinal position, i.e., the bunch length,
as $\sigma_z^i$,
and the RMS spread in the energy as $\sigma_{{\Delta E}/{E}}^i$.
We assume a simple case 
with no correlation between
the longitudinal position and energy.
The energy gain of an RF acceleration field can be written as
\begin{equation}
U = \frac{E_f-E_i}{\cos(\phi_{RF})}\cos(\phi_{RF} - 2\pi \frac{f_{RF}}{c} z) ,
\end{equation}
where $f_{RF}$ is the frequency of the RF,
and $\phi_{RF}$ is the acceleration phase relative to the maximum acceleration phase.
$z$ is the longitudinal position of the bunch
with respect to the bunch center.
A positive/negative $z$ denotes the head/tail of the bunch.
We consider the average energy gain of the RF field to be fixed
while $\phi_{RF}$ is changed.
The phase space after the acceleration is shown in (b).
The energy spread after the acceleration is
\begin{equation}
\sigma_{\Delta E/E}^f = 
\frac{2 \pi f_{RF}}{c}
\sigma_z^i \frac{E_f-E_i}{E_f} \tan (\phi_{RF}) \quad .
\end{equation}
Considering the conservation of the volume in the phase space,
the possible minimum bunch length
having the energy spread will be
\begin{equation}
\sigma_z^f = \frac{ E_i \sigma_z^i \sigma_{\Delta E/E}^i}{E_f \sigma_{\Delta E/E}^f}
=
\frac{c}{2 \pi f_{RF}} \frac{1}{\tan(\phi_{RF})} \frac{E_i}{E_f-E_i} \sigma_{\Delta E/E}^i
\quad .
\label{eq:minimumbunch}
\end{equation}

A section of the beam transport line including beam deflection 
and transverse dispersion
has a longitudinal dispersion;
i.e.,
the path length difference depends on the beam energy.
The dependence is described as
$
\Delta z = R_{56} \frac{\Delta E}{E} + R_{566} \left( \frac{\Delta E}{E}\right)^2 + \cdots .
$
$R_{56}$ is the coefficient of the linear term,
and it can be calculated by
\begin{equation}
R_{56} = \int \frac{\eta(s)}{R(s)} ds \quad .
\end{equation}
$s$ is the coordinate along the designed beam path,
$\eta$ is the transverse dispersion,
and $R$ is the curvature radius of the beam path.

When the beam line following the acceleration
is adjusted to be the following value of $R_{56}$,
the minimum bunch length in Eq. \ref{eq:minimumbunch} can be obtained,
corresponding to Fig.\ref{fig:phasespace}(c):
\begin{equation}
R_{56} = - \frac{c}{2 \pi f_{RF}} \frac{1}{\tan(\phi_{RF})} \frac{E_f}{E_f-E_i} \quad .
\end{equation}

\subsection{CTR}

When a relativistic charged particle
crosses the boundary between two media,
transition radiation (TR) is emitted in both forward and backward directions \cite{Happek}.
The spectral and spatial radiation energy 
of the far field for the backward TR
when an electron is incident normally on an infinite metal surface
is given by the "Ginzburg-Frank" formula \cite{Landau} as
\begin{equation}
\frac{d^ 2U}{d \omega d\Omega} = 
\frac{e^2}{4 \pi^3 \epsilon_0 c}
\frac{\beta ^2 \sin^2 \theta}{(1-\beta^2 \cos^2 \theta)^2}
\quad,
\label{eq:GF}
\end{equation}
where $\theta$ is the angle against the backward direction,
$\beta$ is the velocity of the electron normalized by the speed of light $c$,
$\epsilon_0$ is the permittivity of vacuum.
The radiation intensity vanishes at $\theta=0$,
and it has the maximum value at $\theta \sim 1/\gamma$,
where $\gamma$ is the Lorentz factor.
As seen in Eq. \ref{eq:GF},
it does not depend on the frequency $\omega$.
Due to this characteristic of 
spectrum uniformity
up to the plasma frequency of the target material,
TR is useful for measuring
the radiation spectrum originating only from the longitudinal bunch shape.

When a bunch consisting of $N$ electrons emits radiation
by TR or any other mechanism of radiation,
the power is 
\begin{equation}
P(\omega) \cong P_0(\omega) (N + N^2 |f(\omega)|^2)
\quad ,
\label{eq:N2}
\end{equation}
where $\omega$ is the angular frequency of the radiation.
$P_0$ is the power emitted in the case of a single electron.
For TR, $P_0$ does not depend on $\omega$.
$f(\omega)$ is called a bunch form factor,
which is basically the Fourier conversion of the bunch shape:
\begin{equation}
f(\omega) =  \int \rho(t) e^{-i\omega t} dt  \propto \tilde{\rho}(\omega) \quad ,
\end{equation}
where $\rho$ is the normalized charge density distribution,
and $\tilde{\rho}$ is its Fourier conversion.
The first and second terms of Eq. \ref{eq:N2}
represent the incoherent and coherent parts of the radiation, respectively.
When the 
bunch length is shorter than the wavelength,
the form factor increases to approach 1,
and the coherent part dominates the power
due to the $N^2$ dependence.
Because the form factor in the THz spectrum component 
is sensitive to the longitudinal bunch structure in the submillimeter scale,
it can serve as a reference to estimate the bunch length of interest to us.

\subsection{Principle of bunch length measurement}
\label{sec:principlebunchlength}

Bunch length measurement
by using an interferometer system of coherent radiation
has been established in many accelerator laboratories \cite{Kung}\cite{Murokh}\cite{Zhang}\cite{Noz}.
The principle of this method is to measure
the radiation spectrum
from the autocorrelation of the radiation amplitude.

From Eq. \ref{eq:N2}, the power spectrum of coherent radiation 
can be written as
$S(\omega) = N^2 |f(\omega)|^2 P_0 $.
Here we consider an interferometer system
that splits the signal in two and recombines them with a time delay of $\tau$.
The recombined amplitude becomes
\begin{eqnarray}
h(\omega) &\propto& \int \rho(t) e^{-i\omega t} dt + \int \rho(t) e^{-i\omega(t+\tau)}dt \\
&\propto& (1+e^{-i\omega \tau}) \tilde{\rho}(\omega) \quad ,
\end{eqnarray}
and the power can be written as
\begin{equation}
|h(\omega)|^2 \propto |\tilde{\rho}(\omega)|^2 (e^{i \omega \tau} + e^{-i \omega \tau}) + const.
\end{equation}
When we detect the signal with a wideband detector
that integrates the power of the whole spectrum,
the signal as a function of the delay is
\begin{eqnarray}
V(\tau) &=& N^2 P_0 \int |h(\omega)|^2 d\omega \\
&\propto& \int \rho(t) \rho(t+\tau) dt + const.
\end{eqnarray}
Ignoring the constant term,
it corresponds to the autocorrelation of the charge distribution.

For example,
when the bunch has a Gaussian shape of RMS length $\sigma_t$ as
\begin{equation}
\rho(t) \propto e^{-\frac{t^2}{2 \sigma_t ^2}} \quad,
\end{equation}
the signal observed at the interferometer is
\begin{equation}
V(\tau) \propto e^{-\frac{\tau^2}{4 \sigma_t ^2}} + const.
\end{equation}
By scanning the delay $\tau$ of the interferometer,
the bunch length can be obtained from the width of the signal shape.

In a practical setup of the interferometer,
there always are
low- and high-frequency cutoffs in the system
due to the 
sensitivity range of the detector and the transferring optics.
A simple way to include the effect of the low-frequency cutoff
is to introduce an empirical function of a Gaussian filter:
\begin{equation}
u(\omega) = 1 - e^{-\xi^2 \omega^2} \quad,
\end{equation}
where $1/\xi$ characterizes the cutoff frequency
\cite{Murokh}\cite{Noz}.
By including the effect of the filter function,
the signal of the interferometer is modified as follows:
\begin{equation}
V(\tau) \propto
\frac{1}{\sqrt{\sigma_t^2}} e^{-\frac{\tau^2}{4 \sigma_t^2}}
- \frac{2}{\sqrt{\sigma_t^2 + \xi^2}} e^{-\frac{\tau^2}{4(\sigma_t^2 + \xi^2)}}
+ \frac{1}{\sqrt{\sigma_t^2 + 2 \xi^2}} e^{-\frac{\tau^2}{4(\sigma_t^2 + 2 \xi^2)}}
\end{equation}

On the other hand,
the high-frequency cutoff 
gives the same effect as the spectrum limit due to the intrinsic bunch length.
It is difficult to separate the effect from the signal.
The system has to have a wider spectrum bandwidth
compared with the bunch length to be measured.

\section{Experimental setup}
\subsection{Accelerator and the bunch compression scheme}

\begin{figure}[htb]
	\begin{center}
	 \includegraphics[width= 0.9\linewidth]{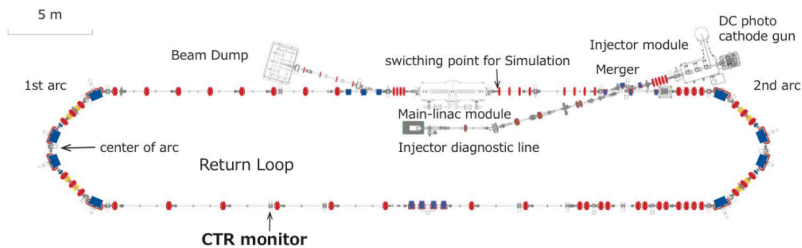}
	  \caption{
Layout of the cERL.
The electron beam provided from the injector 
is accelerated by the main linac,
transferred in the return loop,
then decelerated by the main linac,
and goes to the dump.
}
	\label{fig:cerllayout}
\end{center}
\end{figure}

We briefly describe the accelerator used in our experiment;
a detailed description can be found elsewhere
\cite{Sakanakun}.
Figure \ref{fig:cerllayout} shows the layout of the cERL.
It consists of an injector,
a main linac, and a return loop.
The injector is designed to provide a high average current and a low emittance beam.
It consists of a photocathode DC gun,
a buncher cavity, three sets of two-cell superconducting cavities,
and beam optics matching the downstream optics.
The main linac consists of two sets of nine-cell superconducting cavities in a cryomodule
to accelerate the beam from the injector
and to decelerate it from the return loop
which comes at 180$^{\circ}$ off phase.
Because the RF energy used for the acceleration
is recovered in the deceleration,
the main linac does not experience beam loading in principle,
and it is possible to drive a high current beam.
The return loop is to transport the accelerated beam.
It consists of two arc sections and straight lines.
The beam available at the return loop 
is basically a single-pass beam from the gun. 
Compared with a storage ring system
whose quality is determined by the equilibrium condition of the ring,
the beam quality can be improved with a modern low-emittance photoinjector system.

The nominal beam operation has been performed 
in an on-crest acceleration
with the acceleration phase of the main linac
set to the maximum acceleration.
Because this setting minimizes the energy spread in the return loop,
it has an advantage for beam loss reduction in the beam transport
and stable operation of a high-current beam.
On the other hand, in this study on bunch compression,
the acceleration phase was set to be off-crest,
which increased the energy spread,
and so more careful beam tuning would be required.
We are still on the way in increasing the beam power in the operation.

The machine parameters have been changing
depending on the priority of various study items scheduled in the operational period.
The parameters used in this experiment are summarized in Table \ref{tab:accparam}.
The settings of the injector, especially the phase and amplitude of the three cavities,
were optimized for a bunch charge of 40 pC, 
the typical value in this operational period, 
to realize low emittance and a short pulse
at the input of the main linac.
As the first step of the bunch compression study,
the bunch charge was decreased to 2 pC 
to eliminate complicated collective effects
such as the space charge and the coherent synchrotron radiation kick.
Adding the gun voltage of 400 kV 
and the acceleration in the cavities of 4.2 MV,
the kinetic energy of the injector was obtained as 4.6 MeV.
The two cavities of the main linac can be controlled individually.
The accelerating gradient was not balanced in this operation.
The upstream cavity, ML1,
operated at 5.0 MV/cav.
The RF phase was set for maximizing acceleration, i.e., on the crest,
and  was fixed during this experiment.
The downstream cavity, ML2,
operated at 7.5 MV/cav.
The RF phase was set on the crest initially
and was varied for the bunch compression.
When the phase was varied,
the amplitude of ML2 was adjusted at the same time
to maintain the beam energy.

\begin{table}[htb]
  \begin{center}
    \caption{Beam parameters in this experiment}
    \label{tab:accparam}
    \begin{tabular}{c   c | c } \hline
      Bunch charge & $q$ & 2 pC\\ 
      Kinetic energy at return loop & $E_{loop}$ & 17.1 MeV  \\ 
      Kinetic energy at injector  & $E_{inj}$ & 4.6 MeV \\
      Acc. gradient of upstream cavity & $E_{acc_1}$ & 5.0 MV/cav \\
      Acc. gradient of downstream cavity & $E_{acc_2}$ & 7.5 MV/cav \\
      RF frequency & $f_{RF}$& 1.3 GHz\\
      Rep. frequency of bunches & $f_{b}$& 162.5 MHz\\
      Pulse duration of burst mode & & 0.2 $\mu$s\\
      Pulse repetition of burst mode & & 5 Hz\\
     RMS normalized emittance at injector & $\epsilon_{n}$ & 0.5$\sim$1.5 mm mrad \\
     RMS bunch length at injector (design) & $\sigma_z^i$& $ \sim$0.6 ps \\
      \hline
    \end{tabular}
  \end{center}
\end{table}

The energy recovery is necessary 
for operating the beam at high current continuous mode.
For beam diagnostics, however, 
the beam does not reach the deceleration path
because it is stopped at screen monitors
or the tuning dump at the return loop.
In such cases of energy nonrecovering operation,
the beam power has to be limited
so as not to affect the acceleration gain due to beam loading in the main linac.
We operated in a special mode, burst mode, for beam diagnostics.
In this mode, the beam arrives in a short macro-pulse of the bunch train.
In this experiment, 
the fundamental bunch repetition was 162.5 MHz,
which is 1/8 sub-harmonics of the cavity RF frequency of 1.3 GHz,
and the macro-pulse duration was set to 0.2 $\mu$s.
This macro-pulse repeated at 5 Hz.

\begin{figure}[htb]
	\begin{center}
	 \includegraphics[width= 0.8\linewidth]{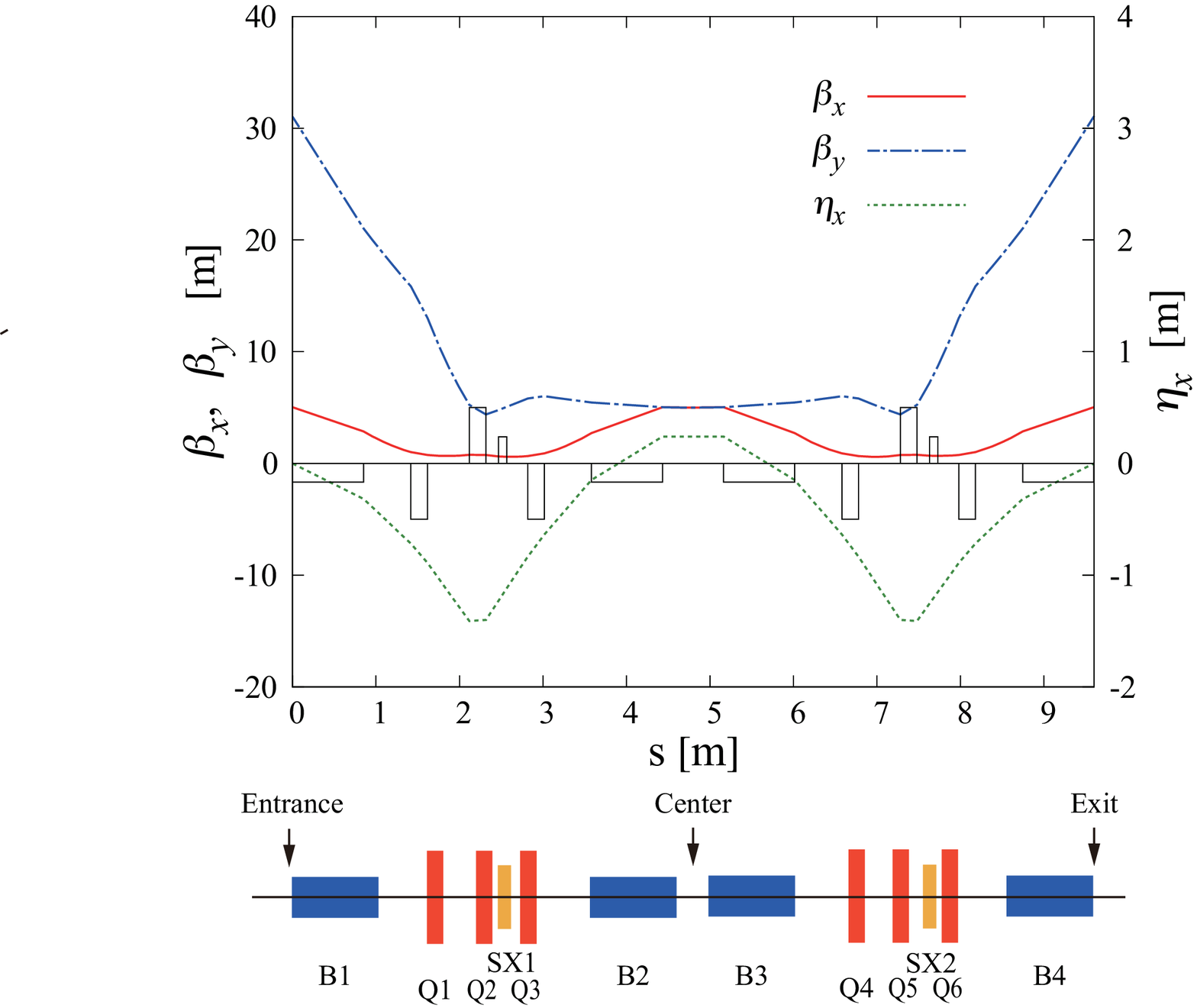}
	  \caption{
Beam optics in the arc section.
The 180$^{\circ}$ arc consists of four bending magnets,
six quadrupole magnets,
and two sextupole magnets.
The optics is designed to be mirror symmetric,
achromat and isochronous. 
}
	\label{fig:arcdesign}
\end{center}
\end{figure}

The short bunch beam at the return loop
is realized by the bunch compression scheme utilizing the first arc section.
The arc section consists of four 45-degrees bending magnets,
six quadrupoles, and two sextupoles.
The beam optics of the arc section is shown in Fig.\ref{fig:arcdesign}.
The linear optics is designed to be mirror symmetric
about the center of the arc.
The nominal optics is designed to satisfy both
the achromat and isochronous ($R_{56}$ to be zero) conditions.
It can also be changed to the non-isochronous condition
while retaining the achromat condition.
By changing the quadrupole magnets in a fixed ratio,
i.e., $\Delta Q1:\Delta Q2:\Delta Q3=\Delta Q6:\Delta Q5:\Delta Q4=1:-2:0$,
the dispersion condition can be tuned while retaining the achromat condition.

In the case of our arc section,
$R_{56}$ has a simple relation 
described only by the dispersion at the center of the arc $\eta_c$.
The calculation of the linear optics in our case gives the relation
$R_{56} = 1.41 \eta_c -0.34$ [m].
Assuming this relation,
$R_{56}$ can be estimated 
from the dispersion measurement of the arc center.

Two sextupole magnets (SX1 and SX2) are installed in the arc section
to correct a nonlinear term of the longitudinal dispersion.
One of the SXs, SX1, which is located between Q2 and Q3 was used 
in this experiment.
The SX was turned off at first
and then scanned 
to experimentally survey the optimum condition for bunch compression.

\subsection{Bunch length monitor}

In order to measure the bunch length 
in the straight section downstream of the arc section,
an extraction port for the CTR was prepared.
In order to minimize the effects of the transverse beam size in the measurement,
the location having a small beam spot size, 
typically $0.4$ mm (RMS) in both planes,
was chosen for the setup, as shown in Fig.\ref{fig:cerllayout}.
An aluminum-coated silicon plate of 70 $\mu$m thickness
can be inserted into the beam line at an angle of 45 degrees with respect to the beam direction.
The backward radiation emitted perpendicular to the beam direction
is extracted to the air through a quartz window of 3.5 mm thickness.
In place of the CTR target,
a Ce-doped YAG scintillator of 100 $\mu$m thickness 
can be inserted.
It is observed by a CCD camera from the other window of the chamber.
This enables us to check the beam profile at the source point of the CTR.

\begin{figure}[htb]
	\begin{center}
	 \includegraphics[width= 0.8\linewidth]{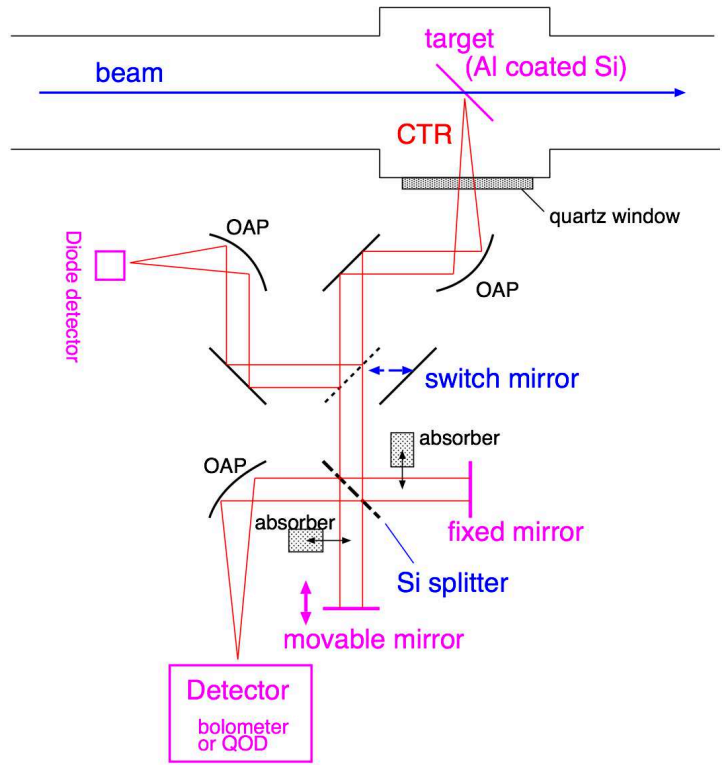}
	  \caption{
Setup of the CTR measurement system.
The source of the radiation is a target inserted in the beam path at 45 degrees.
The system consists of the 
two parts switched by inserting a mirror,
the Michelson interferometer part
and the intensity monitor part. 
}
	\label{fig:setup_interferometer}
\end{center}
\end{figure}

The schematic of the measurement system is shown in Fig.\ref{fig:setup_interferometer}.
The radiation emitted from the target has a diverging angle of $1/\gamma$.
An off-axis parabolic mirror (OAP) of focusing length 225 mm is placed
at the most upstream position in the air to collimate the radiation.
The radiation is then transferred to a Michelson interferometer.
The radiation is split into two paths by 
a high-resistivity silicon plate of 5 mm thickness placed 
at an angle of 45 degrees.
Both paths downstream of the splitter are reflected back with a mirror made of an aluminum plate,
and are recombined at the splitter.
The optical length of the transmitted path can be controlled
by the position of the mirror, which is mounted on a movable stage.
On the other hand, the reflected path of the splitter is fixed.
In order to separately check the alignment and power of each path of the interferometer,
an electromagnetic absorber can be inserted at the middle of each path.
The radiation at the recombined port of the interferometer
is reflected by an OAP.
A THz detector is placed at the focal point.

We used two detectors for the interferometer.
The first one is a compact diode-type detector,
a quasi-optical detector (QOD) that is a product of Virginia Diodes Inc.
Although the sensitivity range of the QOD is specified as 100 GHz to 1 THz,
the spectrum flatness of the sensitivity is poor.
Because it works at room temperature,
it is convenient to quickly check the signal at any time in the beam operation.
Therefore, we used the QOD mainly for the initial setting of the system.
It is mounted on a remote-controlled two-dimensional stage 
for measuring the spatial profile of the radiation 
and for confirming the optimum position to set the detector.
The second detector is a liquid-helium-cooled silicon bolometer,
a product of Infrared Laboratories Inc.
The sensitivity range of the bolometer is 150 GHz to 20 THz 
according to the specification.
The aperture of the bolometer was reduced to 8 mm in diameter
by installing an additional aperture on the input window.
This is for improving the contrast of the interferogram
by improving the spatial resolution in order to observe a fringe pattern.

A simple intensity measurement of CTR 
without an interferometer is useful in beam tuning.
Before entering the interferometer,
a mirror for switching the optical path can be inserted.
The radiation in the switched path is focused by an OAP
and measured by a diode detector.
We used a detector with frequency range 330 GHz to 500 GHz,
WR-2.2ZBD, which is a product of Virginia Diodes Inc,
in this experiment.

\section{Results}

\subsection{Beam optics measurements}

\begin{figure}[htb]
	\begin{center}
	 \includegraphics[width= 0.8\linewidth]{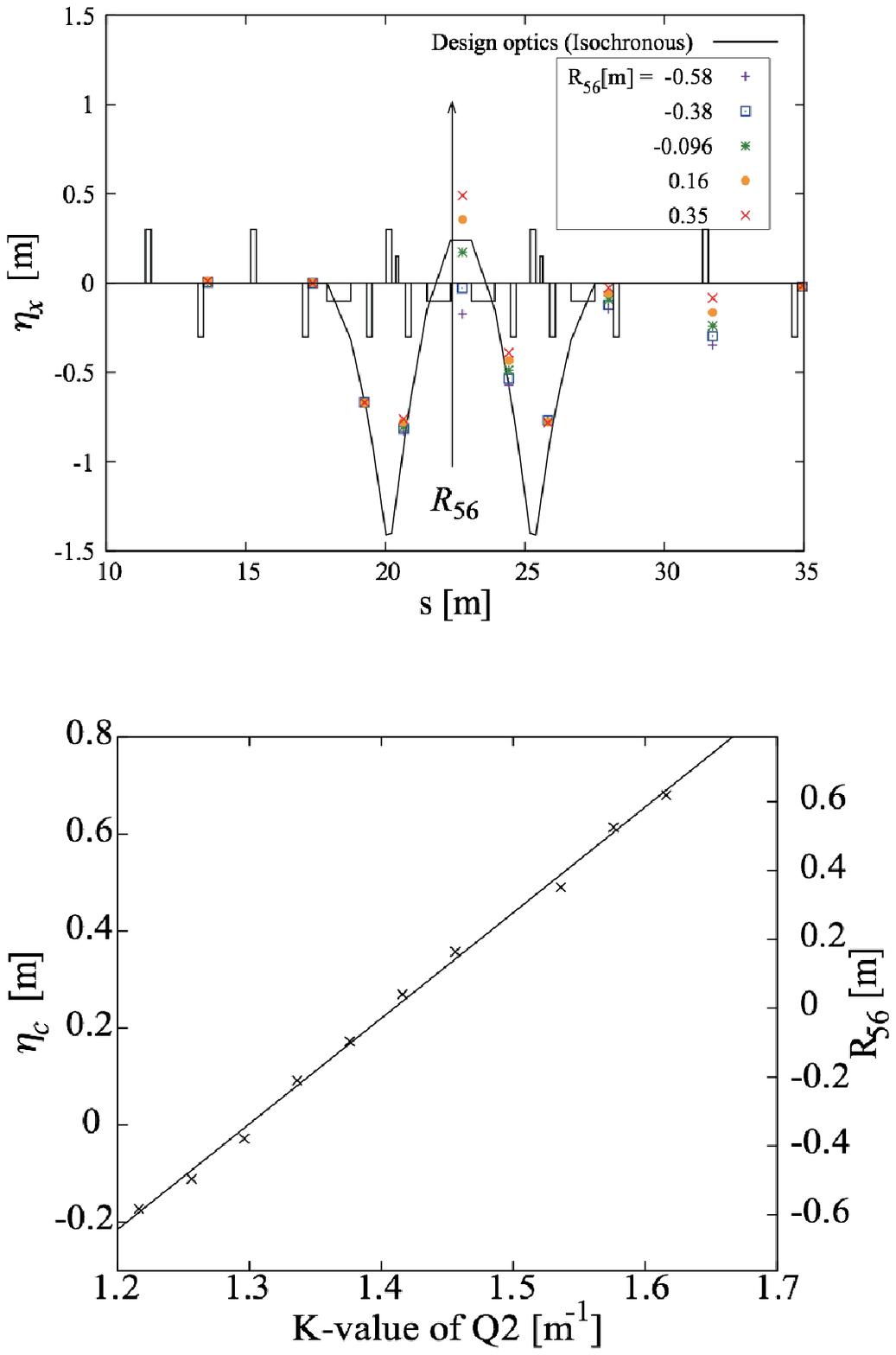}
	 	  \caption{
Measurement of the arc optics.
(Top) 
The horizontal dispersion measured by beam position monitors.
(Bottom)
The relation between the combined knob for $R_{56}$
and the dispersion at the arc center.
The horizontal axis is plotted 
by the strength of one of the quadrupole magnet
used in the combined knob.
}
	\label{fig:opticsmeasurement}
\end{center}
\end{figure}

Beam optics, especially the dispersion function,
of the arc section is critically important for the bunch compression operation.
The dispersion measurement was carried out
by measuring the dependence of the beam trajectory on the beam energy.
The beam energy in the return loop was changed by $-1$\%
by changing the RF amplitude of ML2.
The beam trajectory was measured
using strip-line beam position monitors located along the beam line.
Fig.\ref{fig:opticsmeasurement} (top) shows the measurement results for various settings of $R_{56}$
compared with the designed dispersion function of the isochronous case.
To control $R_{56}$ easily,
a combined knob to change quadrupole magnets at a fixed ratio
was prepared.
This result confirmed that
$R_{56}$ could be controlled by the combined knob
while maintaining the achromat condition
and
the $R_{56}$ of the entire arc section
can be estimated from
the dispersion at the center of the arc.
The relation between 
the dispersion at the arc center $\eta_c$, i.e.,  $R_{56}$
and the settings of the combined knob
is shown in Fig.\ref{fig:opticsmeasurement} (bottom).

\begin{figure}[htb]
	\begin{center}
	 \includegraphics[width= 0.8\linewidth]{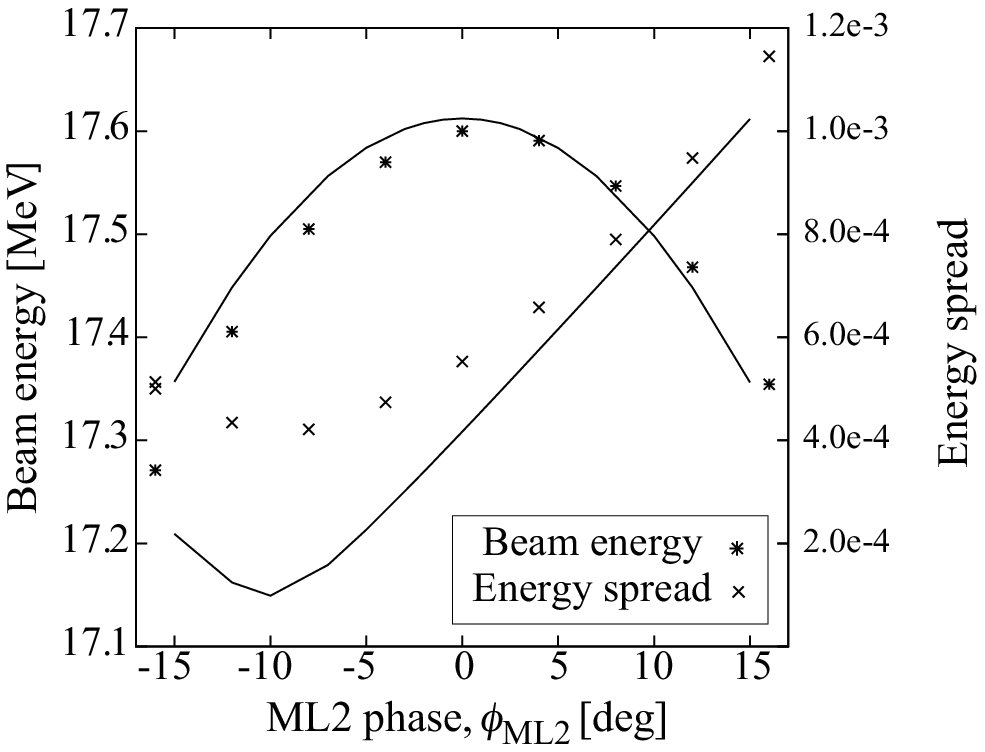}
	  \caption{
Energy and energy spread measurement 
as a function of the ML2 acceleration phase.
The lines show the simulation results.
}
	\label{fig:energymeasurement}
\end{center}
\end{figure}

The acceleration RF phase of the main linac
was confirmed by the beam energy measurement.
The screen monitor located after the first bending magnet
was used for the measurement.
The horizontal position
and the width of the beam profile 
correspond to the beam energy and the energy spread, respectively.
Fig.\ref{fig:energymeasurement} shows the result obtained
by scanning the phase of ML2.
Because the horizontal beam size at the screen monitor
included the contribution of the energy spread and the beta function,
the calculated energy spread could be overestimated 
when the beam size was small.
The measurement result confirmed that 
the phase originated in the maximum acceleration.
It also showed that
the phase having minimum energy spread was shifted about $-10$ degrees
from the on-crest.
This phase shift indicates that the initial beam from the injector
had an energy-time correlation along the bunch.

\subsection{Beam tuning for bunch compression}

\begin{figure}[htb]
	\begin{center}
	 \includegraphics[width= 0.8\linewidth]{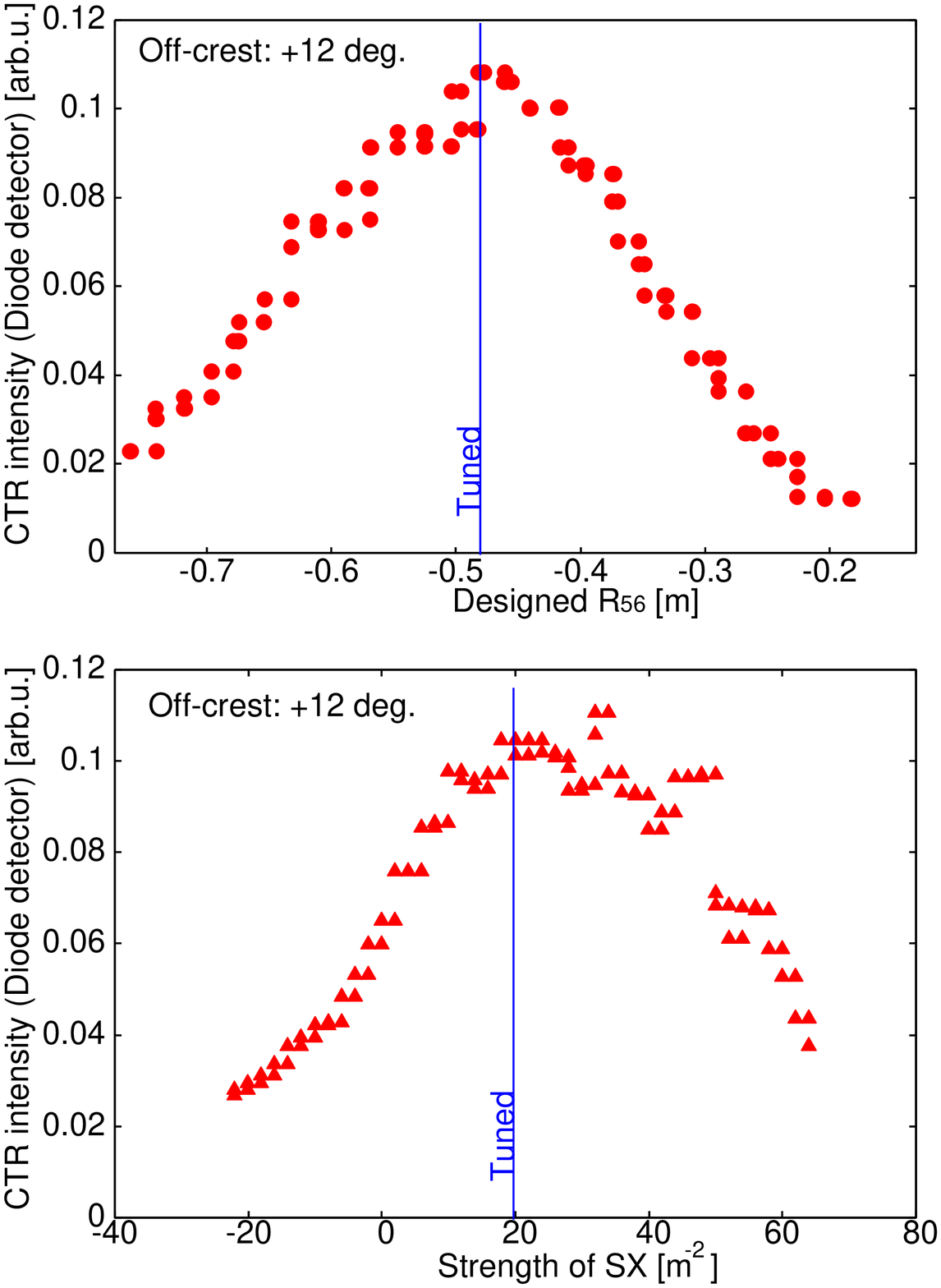}
	  \caption{
Tuning of the arc section for optimizing bunch compression.
(Top) The result of scanning the combined knob of quadrupole magnets.
(Bottom) The result of scanning the sextupole magnet.
}
	\label{fig:arcscan}
\end{center}
\end{figure}

The initial establishment of the beam trajectory in the return loop
was performed in the nominal on-crest mode.
The orbit tuning was accomplished with a beam-based method
using the field center of each quadrupole magnet as a referene.
At first, to switch to the bunch compression mode,
the RF phase of ML2 was shifted to an off-crest value.
The amplitude of ML2 was increased at the same time
to keep the center of the beam energy unchanged.
The beam position at the screen monitor after the first bending magnet
was referred in order to confirm the beam energy.
Next,
the $R_{56}$ of the arc was scanned 
using the combined knob of the quadrupole magnets.
The CTR intensity monitored by a diode detector was used 
as a reference showing the bunch length.
Stronger signals correspond to the shorter bunch.
Fig.\ref{fig:arcscan} (top) shows an example of the $R_{56}$ scan.
The knob was set to the condition giving the maximum.
In order to further correct the effects in higher orders,
the strength of the sextupole magnet was scanned
referring to the CTR strength also as shown in Fig.\ref{fig:arcscan} (bottom).
The scans of $R_{56}$ and the sextupole
were repeated iteratively to find the condition that produced  the strongest CTR.

\subsection{Interferometer measurement and analysis}

After the establishment of the best condition for the bunch compression
by using the CTR intensity measurements,
we carried out the CTR spectrum measurement using the interferometer.
First,
the QOD mounted on a movable stage
was used for checking the optical alignment of the interferometer.
The profile of the radiation at the position of the detector
was checked for each path of the interferometer
by utilizing the absorber to block one of the paths.
The detector was then switched to the bolometer.
The detector signal was recorded 
while changing the optical length of the scanning path in 32 $\mu$m steps. 
To validate the visibility of the interferogram,
measurements for the signal intensity of each single path
and for the signal pedestal by blocking both paths
were also carried out.

\begin{figure}[htb]
	\begin{center}
	 \includegraphics[width= 0.8\linewidth]{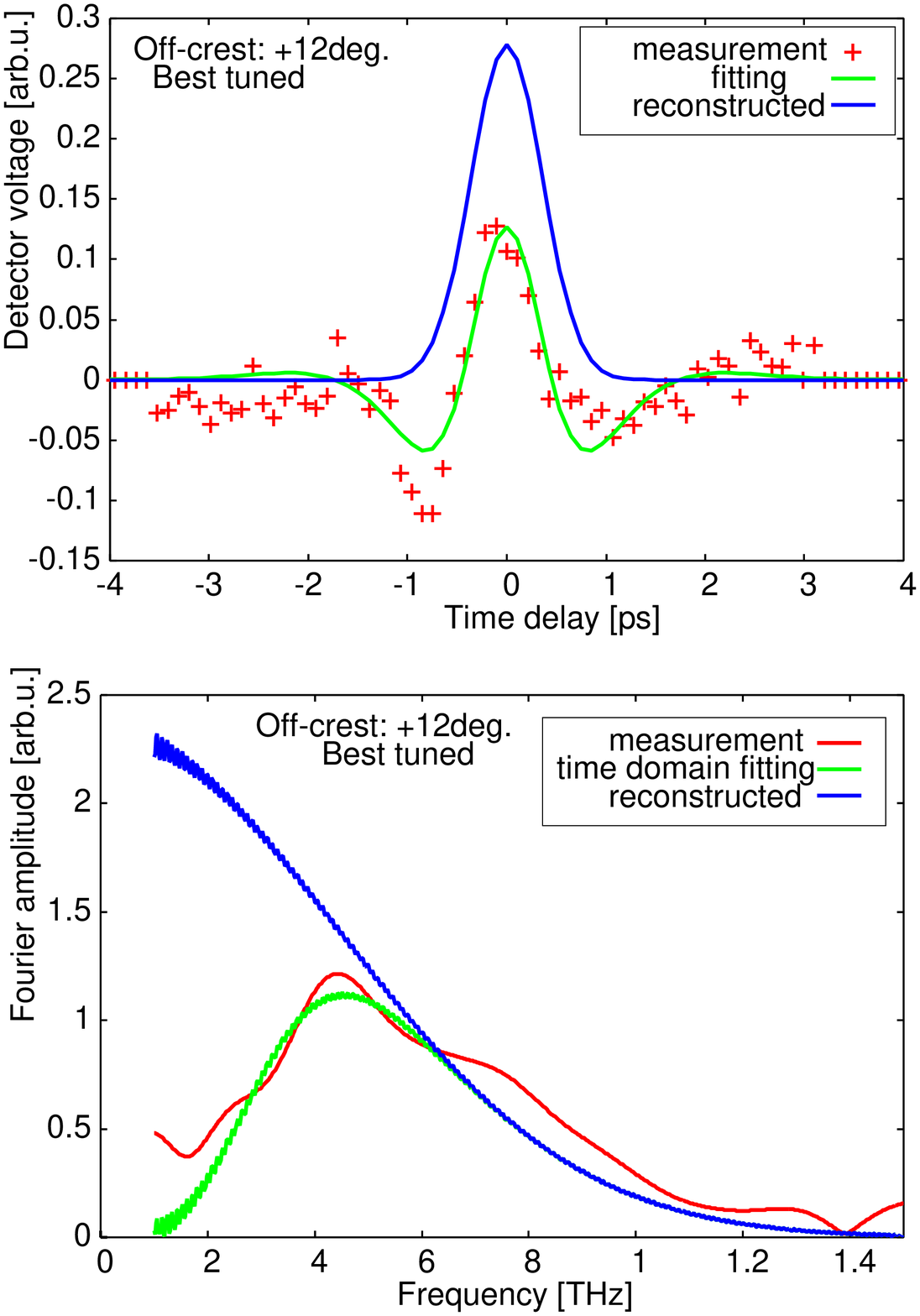}
	  \caption{
Example of data.
(Top) The interferogram and the fitting result with the filter model.
The reconstructed auto-correlation removing the effect of the low-frequency cutoff
is also shown.
(Bottom)
Fourier conversion of the three methods.
}
	\label{fig:datafitting}
\end{center}
\end{figure}

An example of the interferometer signal is shown in Fig.\ref{fig:datafitting} (top).
Because the overall constant does not include information on the bunch length,
the baseline is adjusted to zero.
Each data point shows the average value of multiple measurements
at the same step of the scan.
The data were fitted by a function of the filtered model:
\begin{equation}
V(\tau) =A \left(
\frac{1}{\sqrt{\sigma_t^2}} e^{-\frac{(\tau-\tau_0)^2}{4 \sigma_t^2}}
- \frac{2}{\sqrt{\sigma_t^2 + \xi^2}} e^{-\frac{(\tau-\tau_0)^2}{4(\sigma_t^2 + \xi^2)}}
+ \frac{1}{\sqrt{\sigma_t^2 + 2 \xi^2}} e^{-\frac{(\tau-\tau_0)^2}{4(\sigma_t^2 + 2 \xi^2)}}
\right) + b
\end{equation}
with five free parameters.
$A$ is the overall scale factor, 
$\sigma_t$ is the bunch length, 
$\tau_0$ is the center of the interferogram, 
and $\xi$ is the low-frequency cutoff timescale.
$b$ is the baseline offset.
The fitting result is shown in the plot.
The $\sigma_t$ and its error estimated from the overall fitting
were used as the result of the bunch length.
An asymmetry was found in the interferogram.
The systematic effects due to the disagreements from the model function,
including the effects of the asymmetry, have to be taken care for estimating the error.
An empirical scaling was applied to the statistical error of each data
for the reduced-$\chi^2$ of the fitting to be unity.
This procedure included the systematic error in the results.
The original autocorrelation without the effects of low-frequency cutoff
can be reconstructed 
by replacing $\xi$ with infinity. 
The reconstructed data is also shown in Fig.\ref{fig:datafitting} (top).

The spectrum of the radiation is obtained 
by Fourier transformation of the interferogram.
Fig.\ref{fig:datafitting} (bottom) shows the spectrum calculated in three ways from the same data:
the direct transformation from the data points,
the transformation from the fitted function,
and the transformation from the reconstructed function.
Basically the three approaches agree with each other.

\begin{figure}[htb]
	\begin{center}
	 \includegraphics[width= 0.8\linewidth]{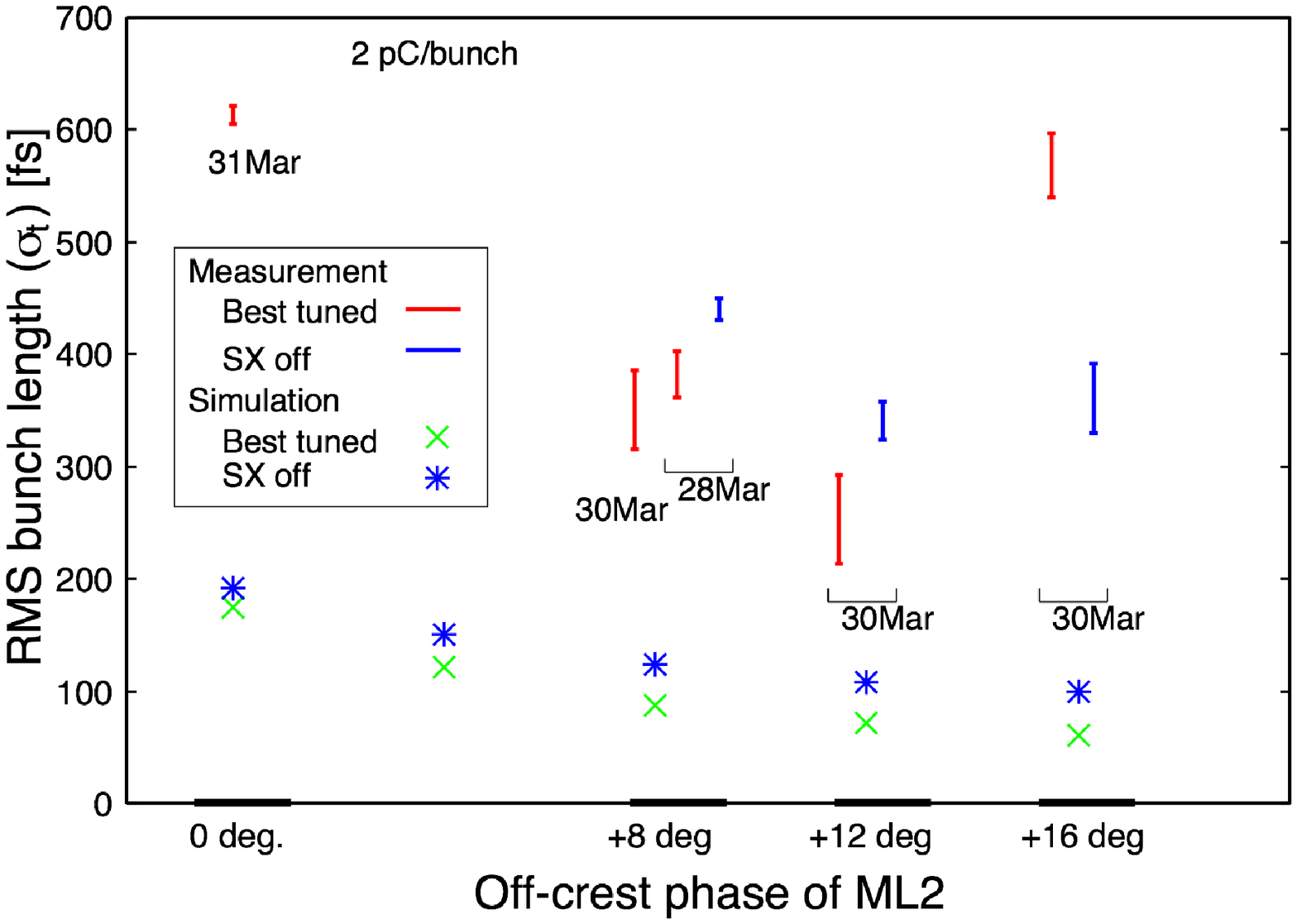}
	  \caption{
Result of the off-crest phase survey.
The best-tuned cases that applied both scans of $R_{56}$ and sextupole magnets
for maximizing the CTR intensity were measured.
Measurements were also taken with 
the sextupole magnet turned off from the best tuned cases (SX off).
}
	\label{fig:offphasescan}
\end{center}
\end{figure}

The bunch compression factor depends 
on the off-crest phase first set to the main linac.
We carried out the tuning and measurement procedure
at several settings of ML2,
namely, at 0, +8, +12, and +16 degrees.
In our definition,
a positive phase means a higher energy gain at the head part of the bunch.
The data were taken on three different days of beam operation,
namely, March 28, 30, and 31 of 2017.
The best tuned cases after applying both the $R_{56}$ scan
and the SX scan were measured.
To check the effect of the sextupole magnet,
measurements were also taken with
the sextupole magnet turned off from the best-tuned cases.
The results are summarized in Fig.\ref{fig:offphasescan}.
The off-crest phase of +12 degrees was found to be the best.
The obtained RMS bunch length was $250\pm50$ fs
at the best condition.

\section{Discussion}

In order to understand the experimental condition,
we performed a particle tracking simulation starting from the gun to the CTR target.
Because the mechanism that dominates the beam dynamics varies depending on the beam energy,
the beam line was divided into two sections,
and different simulation codes were used for each section.
The switching point of the simulation codes
is shown in Fig.\ref{fig:cerllayout}.
The first section,
starting from the gun to the entrance of the main linac
where the beam energy was low 
and thus necessitating a space-charge-dominated calculation,
was calculated using GPT \cite{gpt}.
Transferring the particle data at the output of the first section as the input,
the following section from the main linac to the CTR target
was calculated using ELEGANT \cite{ele}.
ELEGANT does not include space-charge calculation;
instead, it takes much less time for calculating a long beam line than GPT
and is convenient for repeating the calculation to optimize the beam optics.
We note that there still remains 
the effect of the space-charge in our case of the low-energy beam.
In particular, the short section from
the entrance of the main linac module to the location of the cavities
where we did not include the space-charge calculation
might have a non-negligible effect.

The parameters in the injector,
for example,
 the bunch charge, the electron distribution on the gun cathode,
and the amplitudes and phases of the injector cavities
were set according to the experimental conditions.
The phase and amplitude of the main linacs were set
using the same procedure that we employed in the experiment,
and the phase origin was determined 
by finding the maximum acceleration phase.
The combined knob for the $R_{56}$ adjustment
and the sextupole magnet was included in the beam optics
as we did in the experiment.
The particle distribution at the CTR target
was the output of the simulation.
In the bunch compression optimization,
the RMS bunch length at the CTR target was the parameter to minimize.

\begin{figure}[htb]
	\begin{center}
	 \includegraphics[width= 0.8\linewidth]{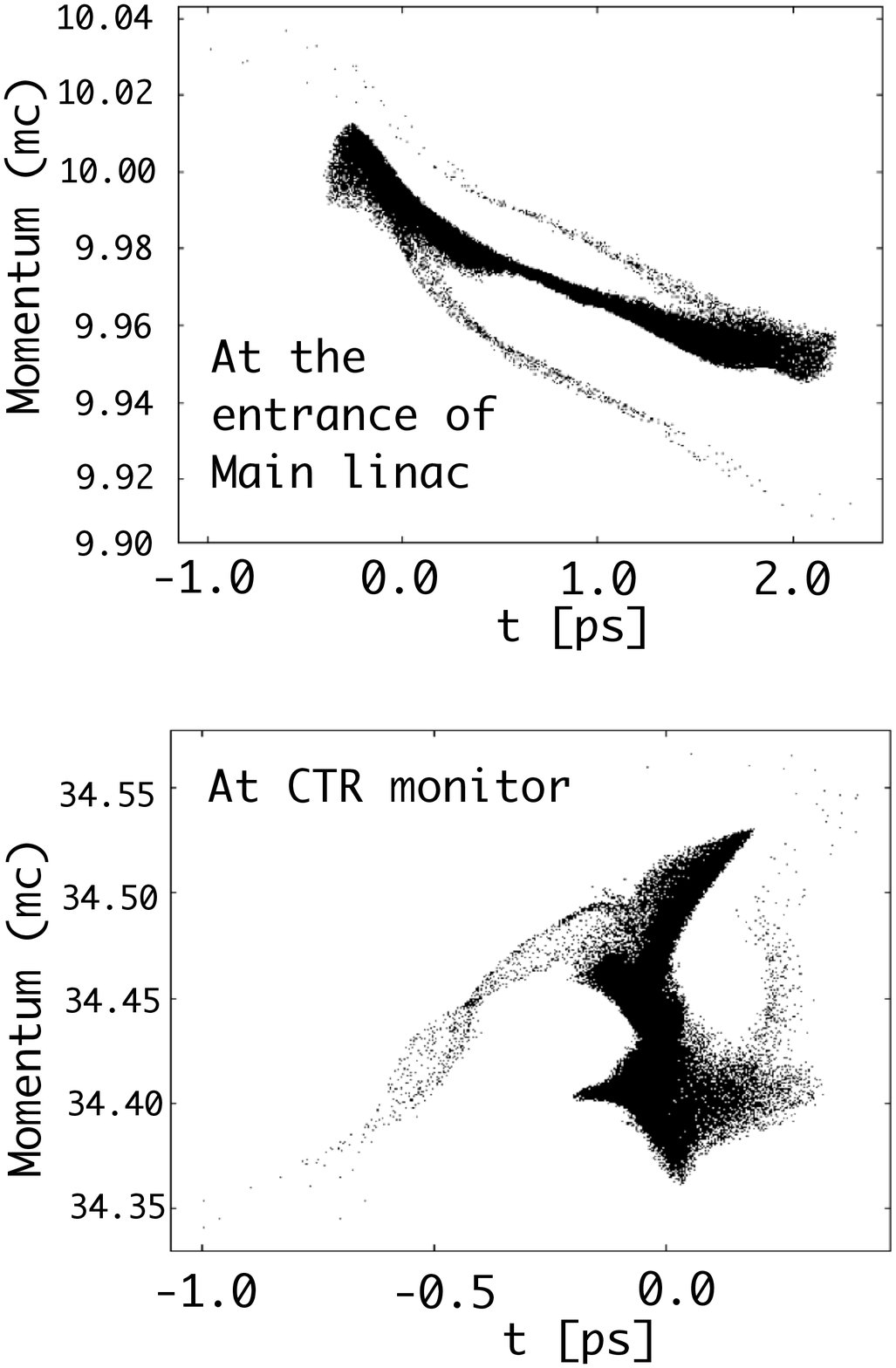}
	  \caption{
Simulated particle distribution in the longitudinal phase space.
(Top) Distribution at the entrance of the main linac 
(switching point of the simulation codes).
(Bottom) Distribution at the CTR monitor.
The best-tuned case for +12 degrees off-crest condition.
}
	\label{fig:simudistribution}
\end{center}
\end{figure}

Figure \ref{fig:simudistribution}(a)
shows the particle distribution in the longitudinal phase space
at the entrance of the main linac.
Although in Sec. \ref{sec:principle}
we described a simple case where
there was no correlation between the energy and time,
the simulated result shows a clear correlation.
In the experimental result shown in Fig.\ref{fig:energymeasurement},
the ML2 phase that gave the minimum energy spread
was shifted from the phase origin by approximately $-10$ degrees.
It corresponds to this energy-time correlation of the initial beam.
Figure \ref{fig:simudistribution}(b)
shows the particle distribution at the CTR target
in the optimum case of bunch compression for the ML2 phase of 12 degrees.
The RMS bunch lengths obtained 
in the simulations by minimizing the bunch length
for each setting of the ML2 phase
are plotted in Fig.\ref{fig:offphasescan}.
The simulation shows a trend:
the bunch length becomes shorter as the off-crest phase increases.
Although there is a discrepancy in the absolute value,
the experimental results show a similar trend.
However,
it seems that the experimental bunch length starts increasing again at 16 degrees.
The reason is not understood at this point.

Although the expected RMS bunch length in the
simulation was 50 fs,
we obtained 250 fs in the experiment.
Here we discuss the possible reasons for this discrepancy.
One of the most difficult problems in the simulation
is the accuracy of the initial conditions.
As shown in Fig.\ref{fig:cerllayout},
there is a straight beam line dedicated to beam diagnostics of the injector.
By measuring the energy spread and the bunch length at this beam line,
the beam characteristics at the entrance of the main linac
can be indirectly confirmed.
The measured energy spread and bunch length agreed well with the simulation.
However, 
what we confirmed were the projections
of Fig.\ref{fig:simudistribution}(a) on the axes.
Because there should be a correlation in the phase space distribution,
there still remain ambiguities in the volume in the longitudinal phase space.
A larger volume in the initial phase space can limit the bunch length in the compression.

In the beam tuning for bunch compression,
we used the CTR intensity monitor, which senses the frequency range 330 to 500 GHz
as the reference.
However, 
for a bunch length shorter than 100 fs,
the most important range should be higher than 1 THz.
If the bunch shape was different from a simple Gaussian,
the spectrum component at a higher frequency could behave differently.
By using a detector for a higher frequency,
the best-tuned condition might be shifted, and 
a shorter bunch length might be measured.
This possibility should be tested in a future beam operation.

We considered the possibility
that the measured bunch length was limited 
by the measurement system.
One important thing to be checked is the effect of the transverse beam size \cite{Lihn}.
Although the spectrum can be determined 
by the three-dimensional shape of the bunch,
we assumed that the effect of the transverse beam size can be neglected.
The typical transverse RMS beam size on the CTR target was 0.4 mm.
In our parameter,
the form factor determined by the transverse beam size
is calculated as 0.98 at 1 THz,
and it decreases to 0.8 at 3 THz.
Although this effect 
can limit the measurement at a bunch length of approximately 50 fs,
the effect on the spectrum obtained in this experiment
seems negligible.

As discussed in Sec.\ref{sec:principlebunchlength},
a high-frequency cutoff in the system can limit the measured bunch length.
Not only the detector sensitivity,
but also the cutoff in the transferring path is important. 
The most questionable element in the system was the vacuum window made of quartz.
The transmission characteristics of this window in THz range was not well understood.
If it has a sharp cutoff at approximately 1 THz,
it can limit the measurement.

This experiment was performed 
at a relatively low bunch charge of 2 pC
as the first step.
For many applications, 
a higher bunch charge of more than 100 pC is required.
Here, 
we note that there are many issues to consider 
in increasing the bunch charge.
A careful management of the space-charge effects becomes important,
especially in the injector section,
and they can affect 
the beam parameter at the input of the bunch compression system.
The charge density increases after the bunch compression,
and the effect of the space-charge may need to be considered
in the return loop, especially for our case of relatively low beam energy.
CSR from the bunch in the bending magnets 
can impart a strong kick on the same bunch
both in the longitudinal and transverse directions.
It can increase the beam emittance and bunch length.
The CSR effects can be severe,
especially for the compact and low-energy case, which is our case.
Continuing the experiments at an increased bunch charge
is critically important for our next step.

Management of beam halo can be an important issue
in a high average current operation.
In particular, in the off-crest acceleration of the bunch compression,
the longitudinal halo can produce a large energy tail
that is difficult to control.
Several ideas have been proposed for halo management.
For example,
improvement of the temporal response of 
the photo cathode of the gun so that the original bunch tail is reduced,
cutting out the halo at the injector with a collimator system,
and
folding the halo with nonlinear beam optics.
Halo management will be tested
when the beam current is increased.

This experiment was performed by stopping the beam at the CTR target
without energy recovery.
To realize a high average current beam,
energy recovery is necessary.
In the bunch compression scheme described in this paper,
the beam in the return loop has a large energy spread
due to the off-crest acceleration.
When decelerating and transporting the beam to the dump,
it is necessary to decrease the energy spread.
The second arc can be used for bunch decompression 
in a similar way as the first arc was used for the bunch compression.
By controlling the $R_{56}$ of the second arc,
the bunch can be decompressed while having an energy chirp.
By decelerating the beam at an off-crest phase,
the beam energy and energy spread can be recovered at the initial condition.
A test of the bunch decompression scheme is planned in the next step.

\section{Conclusion}

A test accelerator for a future large-scale ERL,
cERL has been constructed. 
In order to realize a short bunch beam at the return loop,
an operating mode for bunch compression has been proposed.
By adjusting the non-zero $R_{56}$ in the arc section
for the energy-chirped bunch introduced by the off-crest acceleration of the main linac,
the bunch from the injector can be compressed to sub-picosecond. 
For performing the fine tuning,
a bunch length monitor was necessary at downstream of the arc section.
We have developed a diagnostic system using CTR.
Because the intensity of CTR is sensitive to the bunch length,
we tuned the beam optics in the arc section 
for maximizing the CTR intensity.
By measuring the autocorrelation interferogram of the CTR,
the RMS bunch length was measured to be 250$\pm$50 fs
at the off-crest phase of 12 degrees set for ML2.
Although the achieved bunch length is somewhat larger
compared to that from the simulation,
we confirmed that the tuning procedure worked well.
This test is intended as a first step 
to check the scheme at a low bunch charge of 2 pC without energy recovery.
For the next step,
we plan to increase the bunch charge 
and operate in energy-recovery mode.
The short bunch beam at a high average current
will be useful to produce coherent radiation in the THz spectrum range.
Such a THz source has been proposed as one of the applications of cERL.

\section*{Acknowledgments}
We thank S. Kimura of Osaka university for the help in using the bolometer setup.
We also thank A. Aryshev for advice regarding development of the CTR system.
We appreciate the cERL members for their support
in regard to the beam operation. 
This work was partially supported 
by JSPS KAKENHI Grant Number 16H05991
and
by Photon and Quantum Basic Research Coordinated Development Program 
from the Ministry of Education, Culture, Sports, Science and Technology, Japan.

\section*{References}

\bibliography{mybibfile}

\end{document}